\documentclass[conference,a4paper]{IEEEtran}
\usepackage{multirow}
\usepackage{graphicx}
\usepackage{amsmath}
\usepackage[psamsfonts]{amssymb}
\usepackage{amsxtra}
\usepackage{threeparttable}
\usepackage{url}

\begin{document}

\title{Between Homomorphic Signal Processing and Deep Neural Networks: Constructing Deep Algorithms for Polyphonic Music Transcription}

\author{%
\authorblockN{%
Li Su\authorrefmark{1} 
}
\authorblockA{%
\authorrefmark{1}
Institute of Information Science, Academia Sinica, Taiwan \\
E-mail: lisu@iis.sinica.edu.tw  Tel: +886-2788-3799}
%
%
}

\maketitle
\thispagestyle{empty}

\begin{abstract}
This paper presents a new approach in understanding how deep neural networks (DNNs) work by applying homomorphic signal processing techniques. Focusing on the task of multi-pitch estimation (MPE), this paper demonstrates the equivalence relation between a generalized cepstrum and a DNN in terms of their structures and functionality. Such an equivalence relation, together with pitch perception theories and the recently established rectified-correlations-on-a-sphere (RECOS) filter analysis, provide an alternative way in explaining the role of the nonlinear activation function and the multi-layer structure, both of which exist in a cepstrum and a DNN. To validate the efficacy of this new approach, a new feature designed in the same fashion is proposed for pitch salience function. The new feature outperforms the one-layer spectrum in the MPE task and, as predicted, it addresses the issue of the missing fundamental effect and also achieves better robustness to noise. 
\end{abstract}


\section{Introduction}

Automatic music transcription (AMT) refers to the task of converting acoustic music signals into symbolic notation, such as the onset time, offset time, pitch, and others. Since music is typically composed of multiple overlapped sources with diverse spectral patterns spreading over a wide frequency range, AMT is a highly challenging task, and is considered to be the holy grail in the field in machine listening of music \cite{benetos2012automatic, benetos2013automatic}. The most fundamental technique required for AMT is pitch detection. Depending on the types of source signals and the scenario in application, pitch detection algorithms are designed in different ways, such as single-pitch detection for monophonic music, or melody tracking and and multi-pitch estimation (MPE) for polyphonic music. The main focus of this paper is on the MPE task, the core task in building an AMT system. Solutions to the MPE task include feature-based approaches such as pitch salience functions \cite{yeh2010multiple,su2015combining,su2016exploiting},  nonnegative matrix factorization (NMF) \cite{vincent2010adaptive}, probability latent component analysis (PLCA) \cite{benetos2012shift,benetos2013effcient} as well as convolutional sparse coding (CSC) \cite{cogliati2016context,cogliati2017piano}, to name but a few.

Recently, deep learning approaches have gained increasing attention in polyphonic pitch detection \cite{han2014neural, sigtia2016end, Verma2016FrequencyEF, kelz2016potential,kelz2017experimental}. In \cite{Verma2016FrequencyEF}, Verma and Schafer used time-domain DNNs for melody tracking in polyphonic music, and found that the learned networks resemble traditional pitch detection methods: the first layer behaves like a spectral analyzer while the second layer behaves like a comb filter for saliency mapping. Although such observations keep providing indirect evidence on the physical meanings of deep learning, it is still a highly empirical technique, with its superior performance unexplainable and its limitation unknown. Moreover, unlike its success in other fields such as image processing and speech recognition, deep learning approaches have not been proven state-of-the-art in the problems of pitch detection. In a recent work on MPE, Kelz shows that when the training data contains concurrent notes, neural networks suffer from the \emph{entanglement} issue, which somehow implies a limited deduction power of deep learning on this problem \cite{kelz2017experimental}. These facts reveal the urgency of understanding the theory behind deep learning, and suggest a direction to ask if conventional signal processing techniques can help one understand deep learning.

Recent studies focusing on how to understand deep learning are based on diverse approaches, including scattering transform \cite{bruna2013invariant,mallat2016understanding, wiatowski2015mathematical}, Taylor expansion \cite{montavon2017explaining}, generative modeling \cite{dai2014generative}, renormalization groups \cite{mehta2014exact}, probability theory \cite{patel2015probabilistic}, tensor analysis \cite{cohen2016expressive}, saliency map or layer-wise relevance propagation \cite{bach2015pixel, montavon2017explaining, samek2017evaluating} and cascaded filtering \cite{kuo2016understanding,kuo2017cnn}. Particularly, in \cite{kuo2016understanding,kuo2017cnn}, Kuo interprets the convolutional neural networks (CNN) as the so-called REctified-COrrelations on a Sphere (RECOS) filters, which answer the question why nonlinear activation functions and multi-layer structures are useful: a nonlinear activation function eliminates the information negatively correlated to the anchor vectors (i.e., frequently occurring patterns or dictionary atoms) and, when the filters are cascaded in layers, the system improves itself in each processing step by avoiding confusion with negative-correlation samples, noise, and the background. Since it attempts to bridge the gap between different languages used in signal processing and deep learning, such an approach is noteworthy for the research in audio signal processing among all the above-mentioned approaches.

This paper will show that the structure similar to the RECOS filter can also be applied in MPE and, maybe surprisingly, such a structure do exist in many traditional pitch detection algorithms, including the autocorrelation function (ACF), cepstrum (a.k.a. homomorphic signal processing) \cite{kobayashi1984spectral}, as well as the YIN algorithm \cite{de2002yin}, where the RECOS filters are represented as the Fourier transform followed by a piecewise multiplication. In other words, those traditional pitch detection functions inherently have a DNN-like structure. To the best of the author's knowledge, this fact has never been discussed before. In addition, by reviewing the issues such as the \emph{missing fundamental} effect in pitch perception, this paper will also provide an perceptual interpretation of the nonlinear activation function in a DNN. Moreover, to prove that the proposed statements do inspire one to understand how a deep algorithm works on MPE, a novel feature computed with a three-layer network, named the generalized cepstrum of spectrum (GCoS), is proposed based on the statements. Experiments on piano and mixed-instrument datasets show that the GCoS outperforms the baseline method by showing its ability in detecting missing fundamentals and its robustness to noise. 

This paper is organized as follows. Section \ref{sec: background} reviews two classic topics on the role of nonlinearity in pitch detection, the first being the theories explaining the missing fundamental effects, and the second being the cepstrum. Section \ref{sec:method} gives a generalized framework for pitch detection algorithms and discusses its relation to DNNs and RECOS filters. Remaining parts go for experiments, results and conclusions.

\section{Nonlinearity and pitch detection}
\label{sec: background}
In this section, the nonlinearity of pitch detection will be investigated in the context of perception and signal processing. The missing fundamental effect will be taken as an example of study in the context of psychoacoustics, while cepstrum will be taken as an example in the context of signal processing. In later sections, such investigation will be shown useful to explain the use of nonlinear activation in pitch detection.

\subsection{Missing fundamentals}
The \emph{missing fundamental} effect refers to the phenomenon that one can perceive a pitch when the source signal contains few or no spectral components corresponding to that pitch and its low-order harmonics \cite{rossing2002science}. More specifically, one can perceive the pitch of a note at frequency $f_0$ when there is no spectral component at $f_0$, and even no $2f_0$ in the spectrum -- there could be only spectral components at $nf_0$, $(n+1)f_0,\cdots,(n+k)f_0$ for some $n,k>1$. As a result, pitch detection algorithms cannot be designed by using spectrum only. Though seemingly unusual, the missing fundamental effect does exist commonly in musical signals; it is commonly found in the low-pitch notes of various kinds of instruments such as the 88-key piano, where the physical size of the instrument is too small to effectively radiate the signal with long acoustic wavelength. The same also applies to the music played with a cellphone, where one can still hear bass notes, although distorted, out of a tiny Lo-Fi speaker.

The missing fundamental effect has been studied and explained with mainly three different approaches.\footnote{The discussion here mainly focuses on the main ideas of the theories rather than the relations and debates among them in the history. For more historical backgrounds of the pitch perception theories, readers are referred to \cite{rossing2002science}.} 
First, the period or \emph{time theory} holds that a pitch is determined by the minimal \emph{period} of the waveform rather than the fundamental frequency of the signal. In contrast to the \emph{place theory} which is based on Fourier transform, time theory explains the missing fundamental effect: the period of a signal with spectral components at $nf_0$, $(n+1)f_0,\cdots,(n+k)f_0$ obviously has a period at $1/f_0$. The idea that ``pitch is period'' is the basic assumption of many pitch detection algorithms, including the cepstrum which will be introduced later.

Second, the theories incorporating \emph{nonlinear} effects are many: for example, Helmholtz claimed that one can hear the missing fundamental because of the nonlinear effect in the ear; two sinusoidal components at frequencies $f_1$ and $f_2$ would generate a component at the frequencies $mf_1+nf_2$, $m,n\in\mathbb{Z}$ \cite{von1877sensations}. Although this theory was rejected by Schouten through further experiments \cite{rossing2002science}, it is still inspiring for signal processing: missing fundamentals can be detected by producing cross terms through nonlinear effects. Another example is the \emph{neural cancellation filters} proposed by de Cheveign{\'e} \cite{de1993separation}, which introduces the concept of neural networks for MPE.

Third, the \emph{pattern matching} theory proposed by Goldstein holds that the auditory system resolves the sinusoidal components of a complex sound, and encodes its pitch which best fits the harmonic series of components no matter whether the $f_0$ component exists in the source signal or not \cite{goldstein1973optimum}. The idea of pattern matching is the basis of many MPE algorithms, such as those using NMF or PLCA \cite{vincent2010adaptive, benetos2013automatic}. However, for MPE, a pattern matching approach also faces challenges such as overlapped harmonics and indeterminacy of the lowest note. Since this paper mainly focuses on the role of deep structure and nonlinearity in pitch detection, the pattern matching approach will not be discussed.


\begin{figure}
  \centering
  \includegraphics[width=\columnwidth]{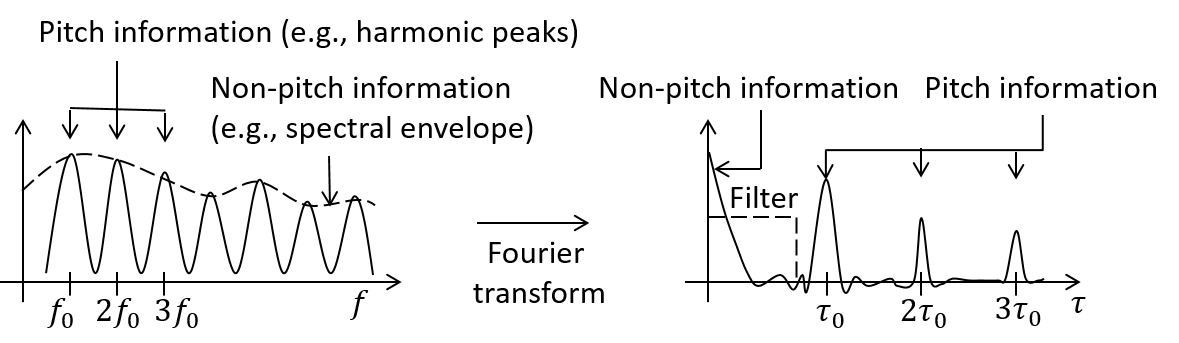}
  \caption{Illustration of a pitch detection algorithm based on the cepstrum.}
  \label{fig: cepstrum}
\end{figure}

\subsection{Cepstrum}\label{sec:cepstrum}

The \emph{cepstrum}, also referred to as \emph{homomorphic signal processing}, is a classic signal processing method with three operations: a Fourier transform, a nonlinear transform (usually a logarithm function), followed by an inverse Foueier transform.
\cite{Oppenheim_Schafer:2009}. Since its invention in 1963 \cite{bogert1963quefrency}, the cepstrum and its derived features have been applied in various signal processing tasks, such as deconvolution, image enhancement, speech recognition, and pitch detection, to name but a few. A thorough review of the cepstrum can be found in \cite{oppenheim2004frequency,Oppenheim_Schafer:2009}.

The cepstrum-based pitch detection algorithm works under a general assumption: 
the fast-varying part of a spectrum, usually the regularly-spaced harmonic peaks, are important for pitch detection, while the slow-varying part, such as the spectral envelope, should be discarded since it is not related to pitch, as shown in Fig. \ref{fig: cepstrum}. To separate the fast-varying part from the slow-varying one, the spectrum is transformed into the cepstrum in the \emph{quefrency} (or \emph{lag}) domain through an inverse Fourier transform.\footnote{Since quefrency and lag have the same unit as time, quefrency, lag and time are used interchangeably in the following discussion.} As a result, the non-pitch information would lie in the low-quefrency (i.e. short-time) region. This term can be discarded by means of a \emph{high-pass filter}. Notice that since the levels of the harmonic peaks in the spectrum vary largely, a nonlinear scaling (usually a logarithm scale) on the spectrum before the inverse Fourier transform is required. Finally, the detected pitch frequency is then the inverse of lag corresponding to the maximal value of the \emph{filtered cepstrum}, as shown in the right part of Fig. 1. The relation between a filtered cepstrum and a DNN will be discussed in the next section.

\begin{figure}
  \centering
  \includegraphics[width=0.9\columnwidth]{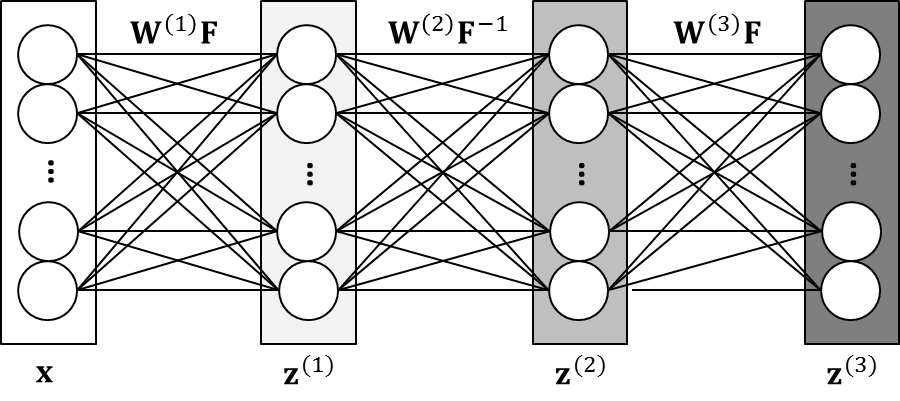}
  \caption{Conceptual illustration of Equations (\ref{eq:power_sepc})-(\ref{eq:2layer_ceps}). Homomorphic signal processing (i.e. cepstrum) is equivalent to a 2-layer network.}
  \label{fig: dnn}
\end{figure}

\section{Generalized Formulation of Pitch Salience Functions}
\label{sec:method}

Since the main focus of this paper is on \emph{frame-level} transcription of polyphonic music, all features are described as the signal processing of an $N$-point, time-domain segment $\mathbf{x} \in \mathbb{R}^N$, a \emph{frame} of music signal in computing the short-time Fourier transform (STFT). All vector throughout this paper are zero-indexed, i.e. $\mathbf{x}=\left[\mathbf{x}[0],\mathbf{x}[1],\cdots,\mathbf{x}[n],\mathbf{x}[N-1]\right]$. The $N$-point windowed discrete Fourier transform (DFT) is denoted by the operator $\mathbf{F}\in\mathbb{C}^{N\times N}$. Similarly, the $N$-point inverse DFT is represented by $\mathbf{F}^{-1}$. Denote $|\mathbf{x}|$ as the absolute value of each element in $\mathbf{x}$.
Consider the following equations:
\begin{align}
\mathbf{z}^{(1)}&=\sigma^{(1)}\left(|\mathbf{F}\mathbf{x}|+\mathbf{b}^{(1)}\right)\,, \label{eq:power_sepc}\\ 
\mathbf{z}^{(2)}&=\sigma^{(2)}\left(\mathbf{W}^{(2)}\mathbf{F}^{-1}\mathbf{z}^{(1)}+\mathbf{b}^{(2)}\right)\,, \label{eq:gen_ceps}\\
\mathbf{z}^{(3)}&=\sigma^{(3)}\left(\mathbf{W}^{(3)}\mathbf{F}\mathbf{z}^{(2)}+\mathbf{b}^{(3)}\right)\,. \label{eq:2layer_ceps}
\end{align}
where $\sigma^{(i)}$ is a element-wise nonlinear transform function such that for $\gamma_i>0$, $i=1,2,3$, 
\begin{equation}
\sigma^{(i)}\left(x\right) = \left\{
  \begin{array}{ll}
    x^{\gamma_i}\,,& \quad x>0\,; \\
    0 \,,& \quad x\leq0\,, \label{eq: general_power}
  \end{array}
  \right.
\end{equation}
$\mathbf{W}^{(i)}\in\mathbb{R}^{N\times N}$ is the weighting matrix for filtering the feature of interest, and $\mathbf{b}^{(i)}$ is the bias vector.

Consider the the first $\lfloor N/2\rfloor$-th elements (i.e. the positive-frequency or the positive-quefrency part) of $\mathbf{z}^{(i)}$. First, assume $\mathbf{W}^{(i)}=\mathbf{I}$ and $\mathbf{b}^{(i)}=\mathbf{0}$ to simplify the discussion. By definition, $\mathbf{z}^{(1)}:=\mathbf{z}^{(1)}[k]$ in (\ref{eq:power_sepc}) is the \emph{magnitude spectrum} of $\mathbf{x}$ estimated from the DFT. The $k$-th element of $\mathbf{z}^{(1)}$, $\mathbf{z}^{(1)}[k]$, is the spectral element at the corresponding frequency $\mathbf{f}[k]=kf_s/N$, $k = 0, 1, \cdots, \lfloor N/2\rfloor$, where $f_s$ is the sampling frequency. Then, $\mathbf{z}^{(2)}$ in (\ref{eq:gen_ceps}) is known as the \emph{generalized autocorrelation function} (GACF) or the \emph{generalized cepstrum} (GC) in the literature, since this formulation is equivalent to the ACF when $\gamma=2$, and is also equivalent to the \emph{cepstrum} up to a linear transformation when $\gamma\to 0$ \cite{kobayashi1984spectral, tokuda1994mel}.\footnote{If the term $x^\gamma$ in (\ref{eq: general_power}) is replaced by $\left(x^{\gamma}-1\right)/\gamma$, then $\sigma_{\gamma,\delta}\left(x\right)|_{x>\delta}$ converges to $\log x$ as $\gamma \to 0$ \cite{kobayashi1984spectral, tokuda1994mel}. In this case, $\mathbf{z}^{(2)}$ is the cepstrum.} As the inverse DFT of a spectral feature in the frequency domain, $\mathbf{z}^{(2)}$ is also said to be a feature in the \emph{lag} or \emph{quefrency} domain, which has the same unit as time. Therefore, for convenience, the elements of $\mathbf{z}^{(2)}$ are indexed by $n$ in this paper. The $n$-th element, $\mathbf{z}^{(2)}[n]$, represents the salience of the feature at the corresponding lag $\mathbf{q}[n] = n/f_s$, $n=0,1,\cdots, \lfloor N/2\rfloor$. The GC has been known as a good feature for MPE by setting $0<\gamma_1<1$, such as $\gamma_1=0.67$ \cite{tolonen2000computationally}, $0.6$ \cite{kraft2014polyphonic}, $0.25$ \cite{indefrey1985design} and $0.1$ \cite{klapuri2008multipitch}. The generalized ACF is also known as  the \emph{root cepstrum}, and it has been shown more robust to noise than the logarithm cepstrum in the literature of speech processing  \cite{lim1979spectral, alexandre1993root}. Interestingly, $\mathbf{z}^{(2)}$ with either $\gamma_1=2$ (i.e. ACF) or $\gamma_1=0$ (i.e. cepstrum) works only for single-pitch detection rather than MPE \cite{rabiner1976comparative, tolonen2000computationally}. One may view GC as a more stable feature than cepstrum since the logarithmic operation is numerically unstable.

Finally, $\mathbf{z}^{(3)}$ in (\ref{eq:2layer_ceps}) is again a feature indexed by $k$ in frequency domain, and $\mathbf{z}^{(3)}[k]$ represents its $k$-th element corresponded to the frequency $\mathbf{f}[k]$. The feature $\mathbf{z}^{(3)}$ in Equation (\ref{eq:2layer_ceps}) is less discussed in the literature of pitch detection. An exception is \cite{peeters2006music}, where Peeters considered using the ACF of the magnitude spectrum as a feature for single-pitch detection, and showed its advantage in identifying the missing fundamentals unseen in a spectrum. This is because that although the peak at the fundamental frequency is attenuated, the ACF still captures the periodicity in between the peaks of high-order harmonics by mixing those peaks up and producing a cross-term at the true fundamental frequency. The ACF of spectrum is a special case of $\mathbf{z}^{(3)}$, where $\gamma_1=2$, $\gamma_2=1$, and $\gamma_3=1$. 
It is therefore straightforward to generalize $\mathbf{z}^{(3)}$ by making both $\gamma_i$ tunable parameters, where $0<\gamma_i\leq 2$. Since $\mathbf{z}^{(3)}$ is computed by a DFT, a nonlinear activation function and another subsequently on $\mathbf{z}^{(1)}$, it can be interpreted as the Generalized Cepstrum of Spectrum (GCoS). This novel feature is firstly proposed and its superior performance to other traditional spectral features will be shown in the experiments. 

As mentioned in Section \ref{sec:cepstrum}, the terms lying in both the low-$k$ and low-$n$ indexes are uninformative for MPE since they are unrelated to pitch. The function of $\mathbf{W}^{(i)}$ and $\mathbf{b}^{(i)}$ is therefore to discard the terms unrelated to pitch or outside the pitch detection region. Assuming that $\mathbf{b}^{(i)}$ is slow-varying, the contribution of $\mathbf{b}^{(i)}$ would simply be the terms discarded by $\mathbf{W}^{(i+1)}$. Therefore, for simplicity, $\mathbf{b}^{(i)}$ is set to be $\mathbf{0}$ throughout this paper, and $\mathbf{W}$ is a diagonal matrix such that
\begin{equation}
\mathbf{W}^{(i)}[l,l] = \left\{
  \begin{array}{ll}
    1 \,,& \quad l>k_c \text{ ($i=1,3$) or } n_c \text{ ($i=2$)}\,; \\
    0 \,,& \quad \text{otherwise}\,. \label{eq: W}
  \end{array}
  \right.
\end{equation}
where $k_c$ and $q_c$ are the indices of the cutoff frequency and cutoff quefrency, respectively.
It means that $\mathbf{W}^{(2)}$ is a high-pass filter with cutoff quefrency at $q_c=n_c/N$ and $\mathbf{W}^{(3)}$ is a high-pass filter with cutoff frequency at $f_c=k_cf_s/N$.\footnote{More precisely, $\mathbf{W}^{(2)}$ should be called a ``long-pass lifter'' rather than ``high-pass filter'' in order to distinguish the filtering processing in the quefrency domain from the one in the frequency domain. This paper uses ``high-pass filter'' for both cases to simplify the terminology.} A straightforward suggestion to the value of $f_c$ and $q_c$ is the lowest pitch frequency and the shortest pitch period. In this paper, $f_c = 27.5$ Hz (frequency of \texttt{A0}) and $q_c = 0.24$ ms (period of \texttt{C8}).

Besides spectrum, cepstrum, and GCoS, there are also other types of pitch salience functions which can be summarized by Equations (\ref{eq:power_sepc})-(\ref{eq:2layer_ceps}). 
For example, the pitch salience function of the YIN algorithm is \cite{de2002yin}:
\begin{equation}
\sum^{N}_{q=1}\left(\mathbf{x}[q]-\mathbf{x}[q+n]\right)^2\,,
\label{eq:yin}
\end{equation}

When the source signal is assumed stationary, derivations show that (\ref{eq:yin}) cab be represented by $\mathbf{z}^{(2)}$, where $\gamma_1=2$, $\gamma_2=1$, $\mathbf{W}^{(2)}=-2\mathbf{I}$, and $\mathbf{b}^{(2)} = 2\mathbf{z}^{(2)}[0]$. This bias term makes the algorithm more stable to amplitude changes.\footnote{Equations (\ref{eq:power_sepc})-(\ref{eq:2layer_ceps}) and the discussion on ACF are based on the assumption that the source signal is stationary. If the source signal is non-stationary, the formulation of (\ref{eq:power_sepc})-(\ref{eq:2layer_ceps}) as well as the resulting weighting and bias factors here should be slightly modified Since the Wiener-Khinchin theorem does not hold anymore. The resulting network should be formulated in a structure like a CNN and will not be discussed here.}

\subsection{Relation to DNNs}

As illustrated in Fig. \ref{fig: dnn}, Equations (\ref{eq:power_sepc})-(\ref{eq:2layer_ceps}) explicitly resemble a DNN with three fully-connected layers. Specifically, a Fourier spectrum is equivalent to a one-layer network, an ACF, cepstrum, or GC are equivalent to a two-layer network, and an ACF of spectrum or GCoS are then equivalent to a three-layer network. 

Equations (\ref{eq:power_sepc})-(\ref{eq:2layer_ceps}) also possess some important characteristics which are different from common DNNs:

\begin{enumerate}
\item The fully-connected layers are complex numbers (i.e., the DFT matrix) while the commonly-used DNNs are typically real-valued. Notice that since $\mathbf{z}^{(1)}$ is symmetric, the Fourier transform in the second and the third layers can be replaced by a real-valued discrete cosine transform (DCT) matrix without changing the result.
\item The nonlinear activation function is a power function; the widely-used rectified linear unit (ReLU) function is a special case where $\gamma=1$. However, previous studies consistently suggest a rather sharp nonlinearity where $\gamma<1$, since taking $\gamma=1$ does not work well in MPE \cite{tolonen2000computationally,kraft2014polyphonic,indefrey1985design,klapuri2008multipitch}.
\item Since the fully-connected layers are determined clearly by the Fourier transform, each feature in each layer has a clear dimensional unit. For example, $\mathbf{z}^{(1)}$ and $\mathbf{z}^{(3)}$ are in the frequency domain while $\mathbf{z}^{(2)}$ is in the time domain. 
\end{enumerate}

Moreover, it should be emphasized that, although the network parameters (i.e., the Fourier coefficients) mentioned here are predetermined rather than learned, they do share similar properties (i.e., frequency selection) in performing pitch detection \cite{Verma2016FrequencyEF}. Therefore, the analogy between a cepstrum and a DNN is not only physically plausible but also provides a new way for one to better understand why and how a DNN works in pitch detection problems. This fact can be seen from the following example.

\begin{figure}
  \centering
  \includegraphics[width=\columnwidth]{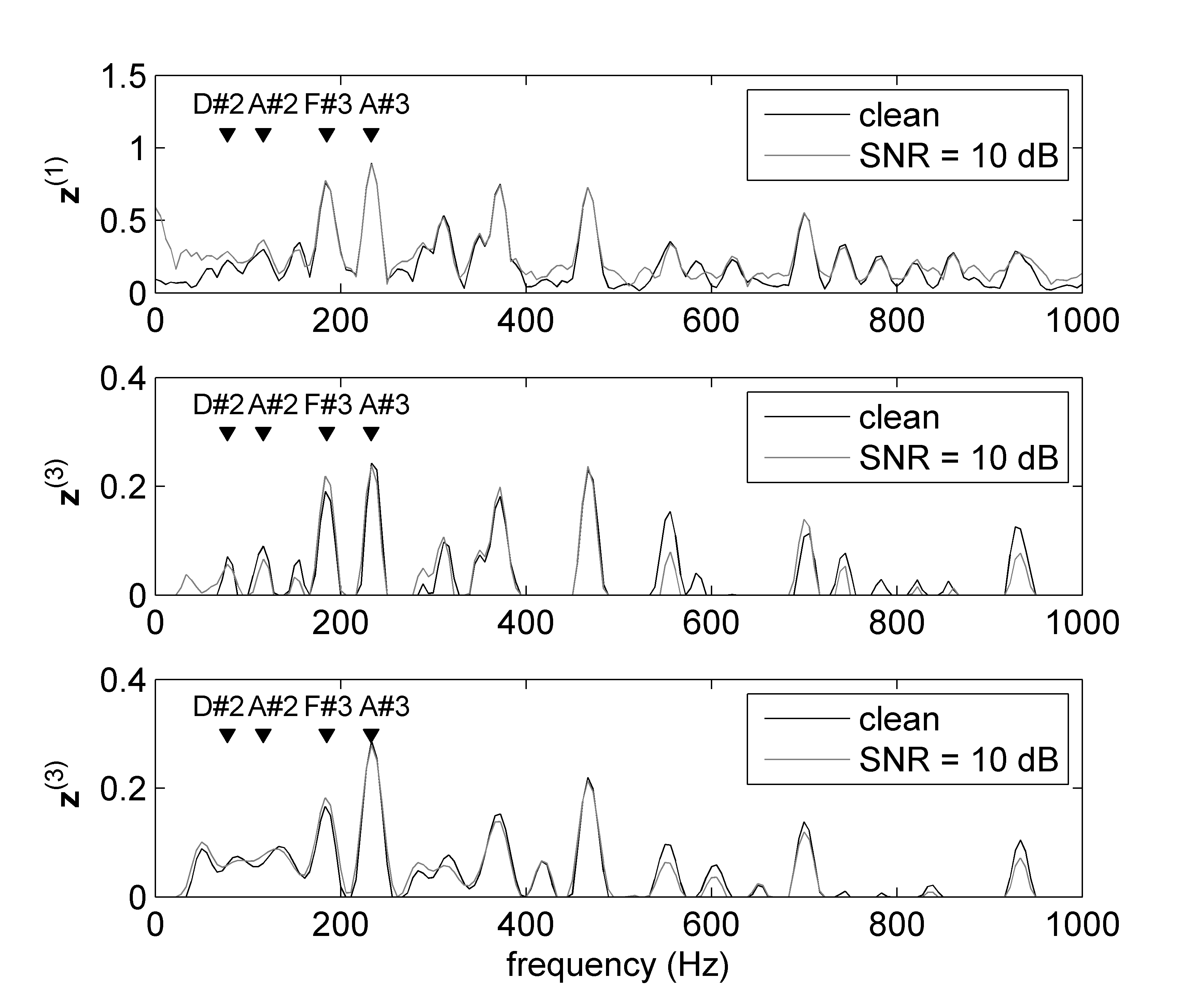}
  \caption{The spectra ($\mathbf{z}^{(1)}$, top), GCoS ($\mathbf{z}^{(3)}$ with $\gamma_1=0.24$ and $\gamma_2=0.6$, middle), and ACF of spectrum ($\mathbf{z}^{(3)}$ with $\gamma_1=2$ and $\gamma_2=1$, bottom) of an audio sample from 17.91 to 18.09 seconds of `MAPS\_MUS-alb\_se2\_ENSTDkCl.wav' (i.e. \emph{Catalu{\~n}a} in Alb{\'e}niz's \emph{Suite Espa{\~n}ola}, Op. 47) from the MAPS dataset. The pitch frequencies of the four notes consisting the sample are labeled.}
  \label{fig: example}
\end{figure}

\subsection{Examples and interpretation}

Fig. \ref{fig: example} shows the spectrum (top) with $\gamma_1=0.24$, GCoS with $\gamma_1=0.24$ and $\gamma_2=0.6$ (middle), and ACF of spectrum (bottom) of a piano signal with four pitches \texttt{D\#2} (77.78 Hz), \texttt{A\#2} (116.54 Hz), \texttt{F\#3} (185.00 Hz), and \texttt{A\#3} (233.08 Hz). In the last two cases $\gamma_3$ is set to 1.\footnote{A different value of $\gamma_3$ merely changes the scale of the output saliency function but does not change the result of pitch detection. Therefore, for simplicity, $\gamma_3$ is set to be 1 for all cases in this paper.} The sampling frequency is $f_s = 44100$ Hz, and the window function $h$ is the Blackman window. To investigate the effect of noise interference on the features, a contaminated signal with pink noise such that the SNR = 10 dB is also considered. All of the illustrated spectra and GCoS features are normalized to unit $l_2$-norm.

The clean and noisy spectra both show that the fundamental frequencies of \texttt{D\#2} and \texttt{A\#2} are weak comparing to their high-order harmonics. There are numbers of peaks unrelated to the true fundamental frequencies and their harmonics, especially in the noisy spectrum. There is also no clear trend in the spectra such that one can set a threshold function to discard those peaks. As a result, the spectrum is not an effective pitch salience function, since it is sensitive to noise and is unable to identify weak fundamental frequencies.

These problems are well resolved in the GCoS. As shown in the middle of Fig. \ref{fig: example}. By setting $\gamma_1=0.24$ and $\gamma_2=0.6$, the GCoS feature enhances two identifiable peaks at the frequencies of \texttt{D\#2} and \texttt{A\#2}. For the noisy signal, most of the fluctuation peaks are eliminated except for a low-frequency peak at 35 Hz, which is the only unwanted cross term produced by the nonlinear activation function. Moreover, the GCoS of the noisy signal is nearly identical to the one of the clean signal; such robustness to noise cannot be seen in spectrum.  

The reasons why GCoS works well are explained as follows. From the discussion in Section \ref{sec:cepstrum}, pitch-related information is the periodic components in the signal. Moreover, the Fourier transform of a periodic signal typically has a harmonic pattern, which is also periodic, while the non-periodic components would lie in the low-frequency or low-quefrency regions. Therefore, the non-periodic components in the $i$-th layer are eliminated by the high-pass filter $\mathbf{W}^{(i+1)}$ in the $(i+1)$-th step. The periodic components are preserved and enhanced through the Fourier transform $\mathbf{F}$ in each step. Those components having negative correlation to the DFT basis, mostly noise and non-periodic ones, are eliminated through rectification in $\sigma^{(i)}$. In other words, similar to the discussion in \cite{kuo2016understanding,kuo2017cnn}, $\mathbf{W}^{(i)}\mathbf{F}$ is a set of RECOS filters for pitch detection, where its anchor vectors form the Fourier basis. 

For the ACF of spectrum, there could also be an alternative interpretation: since $\mathbf{z}^{(3)}$ is an ACF, the anchor vector could be the time shifting of the input spectrum (i.e. $\mathbf{z}^{(1)}$) itself. It computes the correlation of the spectrum to the time shifting of the itself and discards the components inducing negative correlation. Such a viewpoint also explains the reason why the ACF of spectrum is useful in single-pitch detection \cite{peeters2006music}. 


It should be noticed that, however, the ACF of spectrum is not feasible for MPE. The bottom of Fig. \ref{fig: example} shows that, although there are some cross terms produced in the ACF of spectrum, the peaks of these cross terms are actually not at the frequencies of \texttt{D\#2} and \texttt{A\#2}, possibly because of other nonlinear effects produced by other terms. This is the reason why a sharp nonlinear activation function, i.e. $0<\gamma_i\leq 1$, is usually suggested for MPE.

\subsection{Combining features in different layers}

The idea of combining frequency-domain and time-domain features has been proposed by Peeters \cite{peeters2006music} and Emiya \emph{et al.} \cite{emiya2007parametric} for single-pitch detection, and by Su \emph{et al.} for MPE. The idea works well in detecting the pitch frequencies by utilizing the complementary structure of the two features: the frequency-domain features with harmonic peaks and the time-domain features with sub-harmonic peaks. The frequency-domain feature is prone to upper octave errors while robust to lower octave errors \cite{lee2012multipitch,lisu2014smc}, and the time-domain feature be vice versa.\footnote{Octave error means the estimated pitch is off by one or multiples of octave.} Therefore, combining them together could suppress both the upper and the lower octave errors. In other words, fusion of two features at different layers, one in the time domain and the other in the frequency domain, gives rise to an improved pitch salience function $L:=L\left(\mathbf{z}^{(i)},\mathbf{z}^{(i+1)}\right)$. More specifically, we consider
\begin{equation}
L\left(\mathbf{z}^{(1\text{ or }3)},\mathbf{z}^{(2)}\right)=\mathbf{z}^{(1\text{ or }3)}[k]\mathbf{z}^{(2)}\left[\lfloor\frac{N}{k}\rceil\right]
\end{equation}
where the time-domain feature $\mathbf{z}^{(2)}$ is nonlinearly mapped into the frequency domain, and $\lceil\cdot\rfloor$ is the rounding function. For example, \cite{peeters2006music} considered some cases including the combination of the ACF of spectrum ($\mathbf{z}^{(3)}$, $\gamma_1=1$, $\gamma_2=2$, $\gamma_3=1$) and cepstrum ($\mathbf{z}^{(2)}$, $\gamma_1=0$, $\gamma_2=1$).
In this paper, $\mathbf{z}^{(2)}$ and $\mathbf{z}^{(3)}$ are constructed with the same set of networks, while the parameters $\gamma^{(i)}$ are arbitrary values between 0 and 2. Besides, to improve the performance, more steps for post-processing and pitch selection are used.


\section{Experiment settings}

Since this paper is to investigate the multi-layer construction of pitch-related features, a learning-based framework for MPE is not considered now. the MPE method combining $\mathbf{z}^{(1)}$ and $\mathbf{z}^{(2)}$ introduced in \cite{su2015combining} is taken as the baseline method for comparison. 
The purpose of the experiment is to compare two frequency-domain features, GCoS ($\mathbf{z}^{(3)}$) and magnitude spectrum ($\mathbf{z}^{(1)}$), and show that $\mathbf{z}^{(3)}$ outperforms $\mathbf{z}^{(1)}$ in the following senses:
\begin{enumerate}
\item elegant design: the feature can be used directly without pseudo whitening and adaptive thresholding like in \cite{klapuri2003multiple, su2015combining} to filter out the spectral envelope in $\mathbf{z}^{(1)}$ while achieve a more succinct feature representation.
\item detect missing fundamentals: the method can be designed for detecting missing fundamentals without hand-crafted rules in \cite{su2015combining}.
\item robustness to noise: $\mathbf{z}^{(3)}$ outperforms $\mathbf{z}^{(1)}$ because it cancels the uncorrelated parts with more layers. This effect could be more obvious with low SNR.
\end{enumerate}

The source code of the proposed as well as the baseline methods will be announced publicly.
\subsection{Pre-processing}
\label{sec:prep}



After computing the frequency-domain feature and the time-domain feature, the same procedure in \cite{su2015combining} is adopted to compute the pitch profiles of the both features for pitch selection. The features pairs $\left(\mathbf{z}^{(3)},\mathbf{z}^{(2)}\right)$ are mapped into the pitch profiles $\left(\bar{\mathbf{z}}^{(3)},\bar{\mathbf{z}}^{(2)}\right)$ through a filterbank which center frequencies are according to the equal-tempered scale indexed by the \emph{pitch number} $p$ such that $p=\mathcal{P}(f)=69+\lfloor12\log_2 f/440\rceil$,
where $\lfloor\cdot\rceil$ denotes the round function. For \texttt{A0} ($f_0=440$ Hz), $p=69$. Then, the elements corresponding to the pitch number $p$ in $\mathbf{z}^{(3)}$ and $\mathbf{z}^{(2)}$ are merged into the $p$-th element respectively in $\bar{\mathbf{z}}^{(3)}$ and $\bar{\mathbf{z}}^{(2)}$ through max pooling. That is,
\begin{eqnarray}
\bar{\mathbf{z}}^{(3)}[p]&=&\max_{k^\prime} \mathbf{z}^{(3)}[k^\prime] \quad\mathrm{s.t.}\quad \,\,\mathcal{P}\left(f_s\frac{k^\prime}{N}\right)=p\,, \\
\bar{\mathbf{z}}^{(2)}[p]&=&\max_{n^\prime} \mathbf{z}^{(2)}[n^\prime] \quad\mathrm{s.t.}\quad \,\,\mathcal{P}\left(\frac{f_s}{n^\prime}\right)=p\,.
\end{eqnarray}

It is worth mentioning that this process resembles a max-pooling layer in a CNN. However, to make the feature fit the log-frequency scale for pitch detection, the filter size and the stride vary with the frequency and the time index.

\subsection{Pitch selection process}

The criteria of selecting the pitch and the sparsity constraints for reducing false positives proposed in \cite{su2015combining} are applied both the proposed and baseline methods. Specifically, by setting the harmonic/subharmonic series are with length four (i.e. the fundamental frequency/period and the first three harmonics/subharmonics) and the sparse constraint parameter $\delta=0.8$, a pitch $p_i$ is a true positive if 

\begin{enumerate}
\item $\bar{\mathbf{z}}^{(3)}[p_i], \bar{\mathbf{z}}^{(3)}[p_i+12], \bar{\mathbf{z}}^{(3)}[p_i+19], \bar{\mathbf{z}}^{(3)}[p_i+24] > 0$
\item $\bar{\mathbf{z}}^{(2)}[p_i], \bar{\mathbf{z}}^{(2)}[p_i-12], \bar{\mathbf{z}}^{(2)}[p_i-19], \bar{\mathbf{z}}^{(2)}[p_i-24] > 0$
\item $\|\bar{\mathbf{z}}^{(3)}[p_i:p_i+24]\|_0<25\delta$ or $\|\bar{\mathbf{z}}^{(2)}[p_i-24:p_i]\|_0<25\delta$
\end{enumerate}
where $p_i:p_i+24$ means the indices from $p_i$ to $p_i+24$. All pitch activations satisfying these conditions form a piano roll. Finally, a median filter with length of 25 frames is applied on the resulting piano roll in order to smooth the note activation and prune the notes with too short duration.

Notice that unlike \cite{su2015combining}, the rules of selecting missing fundamentals and stacked harmonics are not applied in this works. Experiments will show that the proposed method can properly catch the information of missing fundamentals without introducing these rules.

\begin{table}[t]
\begin{center}
\caption{MPE Result on MAPS and TRIOS datasets. Precision (P), recall (R) and F-score (F)  in \% are listed.}
\begin{tabular}{|l|l|ccc|ccc|}
\hline
\multirow{2}{*}{Dataset}&\multirow{2}{*}{Pitch}&\multicolumn{3}{c|}{Proposed} & \multicolumn{3}{c|}{Baseline}  \\
\cline{3-8}
&&	P&	R&	F&	P&	R&	F \\
\hline\hline
&All&	69.91&	68.94&	\textbf{69.42}&	70.51&	68.26&	69.37\\
MAPS&Bass&	64.14&	48.71&	\textbf{55.37}&	86.51&	39.38&	54.13\\
&Treble&	70.57&	72.03&	\textbf{71.29}&	71.14&	70.95&	71.05\\
\hline\hline
&All&	70.57&	62.60&	\textbf{66.34}&	81.67&	53.22&	64.45\\
TRIOS&Bass&	85.99&	54.00&	\textbf{66.34}&	80.64&	54.37&	64.95\\
&Treble&	68.25&	64.55&	\textbf{66.35}&	81.92&	52.96&	64.33\\
\hline
\end{tabular}
\label{tab:result}
\end{center}
\end{table}

\subsection{Datasets}

We consider two MPE datasets in our evaluation. The first dataset, MAPS, is a widely used piano transcription dataset created by Emiya \textit{et al.} \cite{emiya2010multipitch}.\footnote{\url{http://www.tsi.telecom-paristech.fr/aao/en/2010/07/08/}} It contains audio recordings played on Yamaha Disklavier, an automatic accompaniment piano, in accordance with MIDI-aligned ground-truth pitch annotations. Following previous work \cite{vincent2010adaptive,benetos2012shift,benetos2013effcient,lee2012multipitch, keriven2013structured,o2012structured}, we use the first 30 seconds of the 30 music pieces in the subset ENSTDkCl in our evaluation, totalling 15 minutes.

The second dataset, TRIOS, consists of five pieces of fully synthesized music in trios form \cite{fritsch2012high}.\footnote{\url{http://c4dm.eecs.qmul.ac.uk/rdr/handle/123456789/27}}
Each piece contains piano, two other pitched instruments (the two instruments are different for different pieces). One piece has non-pitched percussions. The pitch annotations were for the three pitched instruments in each piece, whose length ranges from 17 to 53 seconds, totalling 3 minutes and 11 seconds.

\subsection{Parameters and evaluation metrics}

The sampling frequency for all source signals is 44.1 kHz. For computation of the frame-level features, the Blackman-Harris window with size of 0.18 sec is used and the hop size is 0.01 sec. The pitch range considered in evaluation is from \texttt{A1} (55.55Hz) to \texttt{C7} (2205 Hz); both the ground truth and the experiment results are restricted in this range. Further, to identify the performance on the missing fundamentals, the \emph{bass} notes, defined as the notes below \texttt{C3}, and the \emph{treble} notes, defined as the notes above \texttt{C3}, are also evaluated separately.

Following the standard in MIREX MF0 challenge,\footnote{\url{http://www.music-ir.org/mirex/wiki/2016:Multiple_Fundamental_Frequency_Estimation_\%26_Tracking}} the performance of MPE is evaluated using the micro-average frame-level Precision (P), Recall (R), and F-score (F). After counting the number of true positives ($N_{tp}$), false positives ($N_{fp}$) and false negatives ($N_{fn}$) over all the frames within a dataset, The evaluation metrics are defined as: $P=N_{tp}/(N_{tp}+N_{fp})$, $R=N_{tp}/(N_{tp}+N_{fn})$ and $F=2PR/(P+R)$.

To evaluate the robustness to noise for both the baseline and proposed methods, we use
the audio degradation toolbox (ADT) \cite{mauch2013audio} to add pink noise with signal-to-noise ratio (SNR) ranging from 30 dB (least noisy) to 0 dB (most noisy) to every piece in the dataset.

\begin{figure}
  \centering
  \includegraphics[width=\columnwidth]{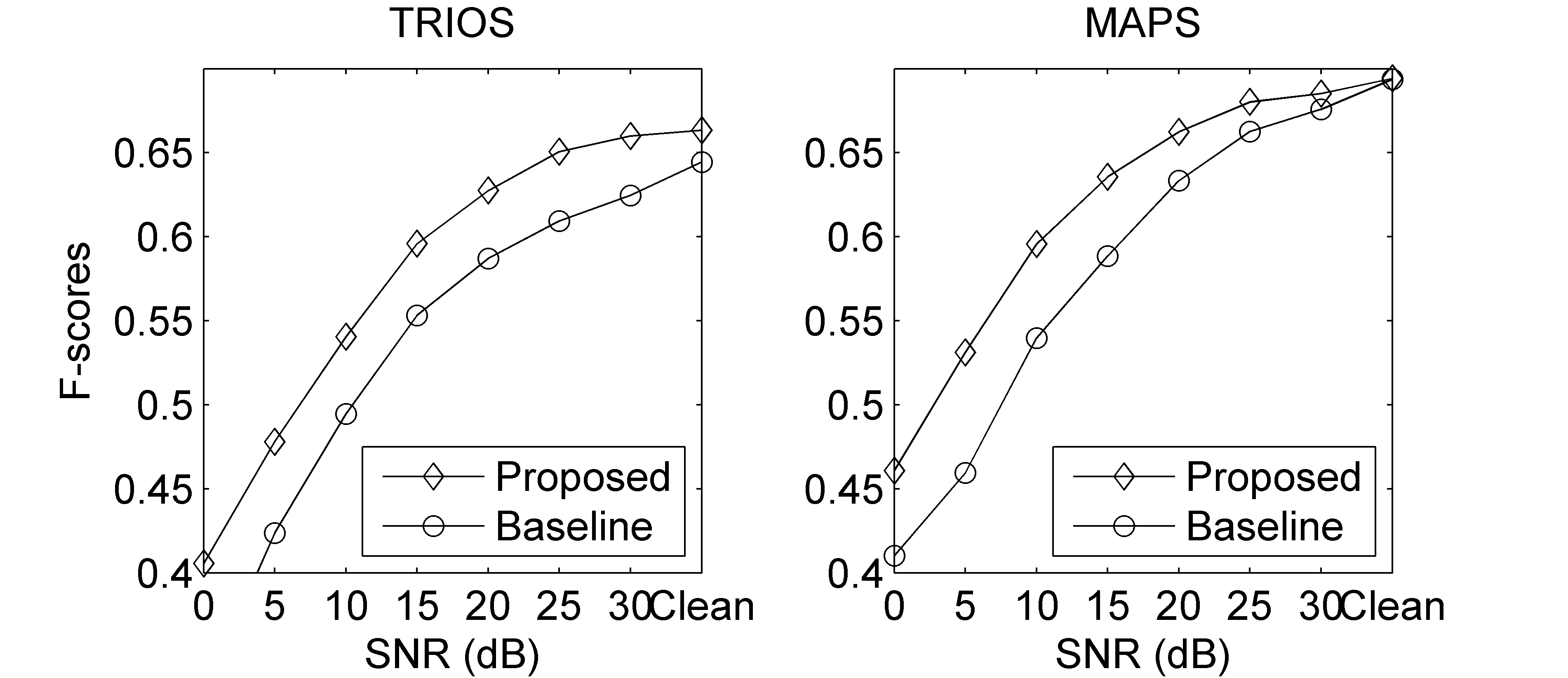}
  \caption{F-scores of the MAPS and TRIOS datasets with different SNR levels for pink noise. ($\circ$): proposed. ($\diamond$): baseline.}
  \label{fig: result}
\end{figure}

\section{Results and discussion}

Table \ref{tab:result} lists the resulting precision, recall and F-score evaluated with 3 sets of pitch ranges: All (\texttt{A1} to \texttt{C7}), Bass, (\texttt{A1} to \texttt{C3}), and Treble (\texttt{C3} to \texttt{C7}). For all pitch sets and datasets, the proposed method outperforms the baseline method in F-scores. This implies an overall improvement of using GCoS.  

For the Bass set, the F-score is improved by 1.24\% in MAPS and by 1.39\% in TRIOS. Such an improvement is more than the ones in the sets of All or Treble in the MAPS dataset. This fact indicates that the proposed feature can better recognize missing fundamentals even without the hand-crafted rules \cite{su2015combining} in selecting missing fundamentals. However, the improvement of the proposed method in either P or R in both datasets behaves inconsistently. In the Bass set of MAPS, the proposed method improves R by 9.33\% while degrades P by 24.37\%. Conversely, in TRIOS, the proposed method improves P while degrades R. The possible reason of such a difference is the property of the input data: TRIOS contains less low-frequency noise since it is constructed with synthetic data. As a result, there are less unwanted low-frequency terms, making more of the cross terms be true pitches, and therefore improving precision more than recall. 



Fig. \ref{fig: result} shows that the improvement of F-score in noisy source data is much more than the improvement in clean data. The lower the SNR is, the more improvement is found in the proposed method. Specifically, when the SNR is lower than 10 dB, the improvement in both datasets is over 5\%. This verifies the statement that the GCoS is more robust to noise since it has one more layer of RECOS filters in refining the features.

\subsection{Discussion}

The idea of using multiple Fourier transforms can be interpreted by one intuition: the Fourier transform of a periodic signal also have periodic (i.e., harmonic) patterns and, perceptually, the strength of such periodic patterns depends on an appropriate nonlinear scaling on the input, and the nonlinear scaling function has some perceptual basis, such as the dB scale or Stevens' power law \cite{stevens1957psychophysical}. This intuition directly suggests promising future directions of applying other kinds of nonlinear activation function and more than three layers of Fourier transform to construct the pitch salience function.

Another remaining issue is that the discussion in this paper still not includes the learning aspect. In fact, the parameter $\gamma_i$ cannot be learned from gradient descent and back propagation, since the gradient of $\sigma^{(i)}$ at zero diverges. Therefore, there is a need to find a differentiable nonlinear activation function which could replace the role of the power function in MPE or, to investigate the potential of other optimization methods such as coordinate descent. Besides, since using the DFT matrix works well in MPE, it is also important to investigate a scenario where a learning-based DNN for the MPE task with the network parameters initialized with a DFT or DCT matrix. 



\section{Conclusions}

This paper has presented a signal-processing perspective for one to better understand the reason why deep learning works by demonstrating a new MPE algorithm which generalizes the concept of homomorphic signal processing and DNN structures. Containing layers of DFT matrices to extract the periodic components, high-pass filters to discard non-periodic components, and nonlinear activation functions to eliminate the components negatively correlated with the DFT matrix, the algorithm has shown superior performance in detecting missing fundamentals and in noisy sources. 
Although the MPE task is merely one specific pattern recognition problem, this approach can still be viewed as an example of RECOS filter analysis for one to better understand deep learning. With the positive experiment results, this paper has positioned one more step to bridge the philosophical and methodological gaps between signal processing and deep learning, and, provides new cues in demystifying and understanding deep learning on other pattern recognition tasks. 





\bibliographystyle{IEEEtran}
\bibliography{mpe_taslp}


\end{document}